\RequirePackage{fix-cm}
\documentclass[twocolumn]{svjour3}          
\smartqed  
\usepackage{graphicx}
\begin{document}

\title{Critical and non-critical coherence lengths in a two-band superconductor
}


\author{Teet \"{O}rd         \and
        K\"{u}llike R\"{a}go      \and
        Artjom Vargunin         
}


\institute{Teet \"{O}rd \and K\"{u}llike R\"{a}go \and Artjom Vargunin \at
            Institute of Physics, University of Tartu \\ T\"{a}he 4, 51010 Tartu, Estonia \\
               \email{teet.ord@ut.ee}          
}

\date{Received: date / Accepted: date}

\maketitle

\begin{abstract}
We study the peculiarities of coherency in a two-gap superconductor. The both intraband couplings, inducing superconductivity in the independent bands, and interband pair-transfer interaction have been taken into account. On the basis of the Ginzburg-Landau equations derived from the Bogoliubov-de Gennes equations and the relevant self-consistency conditions for a two-gap system, we find critical and non-critical coherence lengths in the spatial behaviour of the fluctuations of order parameters. The character of the temperature dependencies of these length scales is determined by the relative contributions from intra- and interband interaction channels.
\keywords{Two-gap superconductivity \and Intra- and interband interactions \and Coherence lengths}
\end{abstract}

\section{Introduction}
\label{intro}

Two-band models of superconductivity have been developed more than fifty years starting from the papers \cite{Ref smw}-\cite{Ref kondo}. For the present the number of discovered multi-gap superconducting materials is quite large, including cuprates, $\mathrm{MgB_{2}}$ and iron-arsenic compounds. In this field various theoretical schemes have been elaborated, see \cite{Ref kw}-\cite{Ref biankoni2} and references therein. In particular, the derivation of high-quality superconducting regions from oxygen ordering, observed recently in La$_{2}$CuO$_{4+y}$ \cite{Ref oxygen}, supports the multi-band theoretical scenario of superconductivity in cuprates.

In a two-gap superconductor with interband pair-transfer interaction one has to deal with coupled band condensates. In this situation the quantities, related initially to the superconducting states of independent bands, acquire mixed character and they describe the collective features of the whole two-component condensate. An example of such quantities in a superconducting system with interband coupling are the coherence lengths which cannot be attributed to different bands involved \cite{Ref babaev}, see also \cite{Ref poluektov}-\cite{Ref kristoffel}. In the present contribution we examine in details the properties of coherence length scales in a two-band superconductor.

\section{System of Ginzburg-Landau equations for a two-band superconductor}
\label{sec:1}
We start with the following Hamiltonian in the terms of $\Psi$-operators for a two-band superconductor with intraband couplings and interband pair-transfer interaction:

\begin{equation}\label{eq1}
H=H_{0}+H_{1} \, ,
\end{equation}
where
\begin{eqnarray}\label{eq2}
H_{0}=\int \mathrm{d}\mathbf{r}\sum_{\alpha }\sum_{s}\Psi^{+}_{\alpha  s}\left(\mathbf{r}\right)
\mathcal{H}_{0}\Psi_{\alpha  s}\left(\mathbf{r}\right) \,
\end{eqnarray}
with
\begin{eqnarray}\label{eq3}
\mathcal{H}_{0}=-\frac{\hbar^{2}}{2m}\left(\nabla - \frac{ie}{\hbar c}\mathbf{A}\right)^{2}-\mu +V(\mathbf{r}) \, ,
\end{eqnarray}
and
\begin{eqnarray}\label{eq4}
H_{1}
&=&\frac{1}{2}\sum_{\alpha ,\alpha'}W_{\alpha \alpha'}\nonumber \\
&\times&\int \mathrm{d}\mathbf{r}\sum_{s, s'}\Psi^{+}_{\alpha  s}\left(\mathbf{r}\right)
\Psi^{+}_{\alpha  s'}\left(\mathbf{r}\right)
\Psi_{\alpha ' s'}\left(\mathbf{r}\right)\Psi_{\alpha ' s}\left(\mathbf{r}\right) \, .
\end{eqnarray}
 In Eqs. (\ref{eq1})-(\ref{eq4}), $\alpha=1,2$ is the band index, $s=\uparrow,\downarrow=\pm$ is the spin index, $\mu$ is the chemical potential, $V(\mathbf{r})$ is the periodic potential of a crystal, and $W_{\alpha \alpha'}$ are the electron-electron interaction constants. We suppose that $W_{\alpha \alpha}<0$, i.e. the autonomous superconducting phase transition takes place in the both bands if interband interaction is absent. However, if interband interaction is turned on, the superconducting states of the bands are not independent anymore.

 The following derivation of the Ginzburg-Landau equations exploits the scheme from \cite{Ref ketterson}, generalized for two-band situation.

 First, one introduces the effective one-electron Hamiltonian in the self-consistent field approximation,
\begin{eqnarray}\label{eq5}
H_{eff}&=& \int  \mathrm{d}\mathbf{r}\left\{\sum_{\alpha }\sum_{s}\Psi^{+}_{\alpha  s}\left(\mathbf{r}\right)
\mathcal{H}_{0}\Psi_{\alpha  s}\left(\mathbf{r}\right)\right. \nonumber \\
&+& \sum_{\alpha ,\alpha'}\sum_{s}U_{\alpha \alpha'}\left(\mathbf{r}\right)\Psi^{+}_{\alpha  s}\left(\mathbf{r}\right)
\Psi_{\alpha ' s}\left(\mathbf{r}\right) \nonumber \\
&+& \sum_{\alpha }\Delta_{\alpha }\left(\mathbf{r}\right)\Psi^{+}_{\alpha  \uparrow}\left(\mathbf{r}\right)
\Psi^{+}_{\alpha  \downarrow}\left(\mathbf{r}\right) \nonumber \\
&+& \left. \sum_{\alpha }\Delta^{\ast}_{\alpha }\left(\mathbf{r}\right)\Psi_{\alpha  \downarrow}\left(\mathbf{r}\right)
\Psi_{\alpha  \uparrow}\left(\mathbf{r}\right) \right \} \, .
\end{eqnarray}
In what follows we simplify the calculations supposing\footnote{The more general situation $U_{\alpha \alpha'}\left(\mathbf{r}\right)\neq 0$ if $\alpha\neq \alpha'$ should be considered as a special problem.}
\begin{equation}\label{eq6}
U_{\alpha \alpha'}\left(\mathbf{r}\right)=U_{\alpha }\left(\mathbf{r}\right)\delta_{\alpha \alpha'} \, .
\end{equation}
In this case the effective Hamiltonian (\ref{eq5}) can be diagonalized by means of the Bogoliubov-Valatin transformation ($\gamma^{+}$ and $\gamma$ are the operators of creation and destruction of elementary excitations):
\begin{equation}\label{eq7}
\Psi_{\alpha  s}\left(\mathbf{r}\right)=\sum_{\mathbf{k}}\left\{u_{\alpha \mathbf{k}}\left(\mathbf{r}\right)\gamma_{\alpha \mathbf{k}s}
-\mathrm{sgn}\left(s\right)v^{\ast}_{\alpha \mathbf{k}}\left(\mathbf{r}\right)\gamma^{+}_{\alpha \mathbf{k}-s}\right\}  \, ,
\end{equation}
where $\mathbf{k}$ is the wave vector. As a result
\begin{equation}\label{eq8}
H_{eff}=E_{g}+ \sum_{\alpha }\sum_{\mathbf{k}}\sum_{s}E_{\alpha }\left(\mathbf{k}\right)
\gamma^{+}_{\alpha \mathbf{k}s}\gamma_{\alpha \mathbf{k}s} \, ,
\end{equation}
where $E_{g}$ is the ground state energy and $E_{\alpha }\left(\mathbf{k}\right)$ is the energy of an elementary excitation. Using the commutator $\left[\Psi_{\alpha  s}\left(\mathbf{r}\right),H_{eff}\right]$ together with Eqs. (\ref{eq7}) and (\ref{eq8}) we obtain the Bogoliubov-de Gennes equations for a two-band superconductor
\begin{eqnarray}\label{eq9}
&&E_{\alpha }\left(\mathbf{k}\right)\left(
\begin{array}{c}
u_{\alpha \mathbf{k}}\left(\mathbf{r}\right) \\v_{\alpha \mathbf{k}}\left(\mathbf{r}\right)\\
\end{array} \right) \nonumber \\
&=&
\left(
\begin{array}{cc}
\mathcal{H}_{0}+U_{\alpha }\left(\mathbf{r}\right)& \Delta_{\alpha }\left(\mathbf{r}\right) \\
\Delta^{\ast}_{\alpha }\left(\mathbf{r}\right)& -\mathcal{H}^{\ast}_{0}-U_{\alpha }\left(\mathbf{r}\right) \\
\end{array} \right)\left(
\begin{array}{c}
u_{\alpha \mathbf{k}}\left(\mathbf{r}\right) \\v_{\alpha \mathbf{k}}\left(\mathbf{r}\right)\\
\end{array} \right) \, .
\end{eqnarray}
Here the self-consistent potentials $\Delta_{\alpha }\left(\mathbf{r}\right)$ and $U_{\alpha }\left(\mathbf{r}\right)$ have been determined as
\begin{eqnarray}\label{eq10}
\Delta_{\alpha}\left(\mathbf{r}\right)&=&\sum_{\alpha '}W_{\alpha \alpha'}\left\langle\Psi_{\alpha ' \downarrow}\left(\mathbf{r}\right)\Psi_{\alpha ' \uparrow}\left(\mathbf{r}\right)\right\rangle \nonumber \\
 &=&-\sum_{\alpha '}W_{\alpha \alpha'}\sum_{\mathbf{k}}v^{\ast}_{\alpha '\mathbf{k}}\left(\mathbf{r}\right)u_{\alpha '\mathbf{k}}\left(\mathbf{r}\right) \nonumber \\
 &\times&\left[1-2f\left(E_{\alpha '}\left(\mathbf{k}\right)\right)\right]\, ,
\end{eqnarray}
\begin{eqnarray}\label{eq11}
U_{\alpha }\left(\mathbf{r}\right)&=&W_{\alpha \alpha}\left\langle\Psi^{+}_{\alpha  \uparrow}\left(\mathbf{r}\right)\Psi_{\alpha  \uparrow}\left(\mathbf{r}\right)\right\rangle \nonumber \\
&=&W_{\alpha \alpha}\left\langle\Psi^{+}_{\alpha  \downarrow}\left(\mathbf{r}\right)\Psi_{\alpha  \downarrow}\left(\mathbf{r}\right)\right\rangle \nonumber \\
 &=&W_{\alpha \alpha}\sum_{\mathbf{k}}\left\{\left|u_{\alpha \mathbf{k}}\left(\mathbf{r}\right)\right|^{2}
 f\left(E_{\alpha }\left(\mathbf{k}\right)\right)\right. \nonumber \\
 &+&\left. \left|v_{\alpha \mathbf{k}}\left(\mathbf{r}\right)\right|^{2}
 \left[1-f\left(E_{\alpha }\left(\mathbf{k}\right)\right)\right]\right\}\, ,
\end{eqnarray}
where $f\left(E\right)=\left[1+\exp\left(E/k_{B}T \right)\right]^{-1}$.

In the spatially homogeneous case Eqs. (\ref{eq10}) and (\ref{eq9}) yield the system of superconductivity gap equations
\begin{eqnarray}\label{eq12}
\Delta_{\alpha }=-\sum_{\alpha '}W_{\alpha \alpha'}\sum_{\mathbf{k}}
\frac{\Delta_{\alpha '}}{2E_{\alpha '}\left(\mathbf{k}\right)}
\tanh\frac{E_{\alpha '}\left(\mathbf{k}\right)}{2k_{B}T}\,
\end{eqnarray}
with $E_{\alpha }\left(\mathbf{k}\right)=\left[\tilde{\varepsilon}^{2}_{\alpha }\left(\mathbf{k}\right)
+\left|\Delta_{\alpha }\right|^{2}\right]^{1/2}$, $\tilde{\varepsilon}_{\alpha }\left(\mathbf{k}\right)=\varepsilon_{\alpha }\left(\mathbf{k}\right)-\mu$, where $\varepsilon_{\alpha }\left(\mathbf{k}\right)$ is the normal-state energy of an electron in the $\alpha$th band.

In the non-homogeneous situation one can derive from Eqs. (\ref{eq10}) and (\ref{eq9}) analogously with \cite{Ref ketterson} the linearized selt-condistency conditions
\begin{eqnarray}\label{eq13}
\Delta_{\alpha }\left(\mathbf{r}\right)=\sum_{\alpha '}\int \mathrm{d}\mathbf{r}'K_{\alpha \alpha'}\left(\mathbf{r},\mathbf{r}'\right)\Delta_{\alpha '}\left(\mathbf{r}'\right) \,
\end{eqnarray}
with
\begin{eqnarray}\label{eq14}
&&K_{\alpha \alpha'}\left(\mathbf{r},\mathbf{r}'\right)=-W_{\alpha \alpha'}\sum_{\mathbf{k},\mathbf{k}'}
\left[1-2f\left(\tilde{\varepsilon}_{\alpha '}\left(\mathbf{k}\right)\right)\right] \nonumber \\
&\times&\left\{\frac{\Theta\left(-\tilde{\varepsilon}_{\alpha '}\left(\mathbf{k}\right)\right)}
{\left|\tilde{\varepsilon}_{\alpha '}\left(\mathbf{k}\right)\right|-\tilde{\varepsilon}_{\alpha '}\left(\mathbf{k}'\right)}+
\frac{\Theta\left(\tilde{\varepsilon}_{\alpha '}\left(\mathbf{k}\right)\right)}
{\left|\tilde{\varepsilon}_{\alpha '}\left(\mathbf{k}\right)\right|+\tilde{\varepsilon}_{\alpha '}\left(\mathbf{k}'\right)}\right\}
\nonumber \\
&\times& \Phi^{\ast}_{\alpha '\mathbf{k}}\left(\mathbf{r}'\right)\Phi^{\ast}_{\alpha '\mathbf{k}'}\left(\mathbf{r}'\right)
\Phi_{\alpha '\mathbf{k}}\left(\mathbf{r}\right)\Phi_{\alpha '\mathbf{k}'}\left(\mathbf{r}\right) \, ,
\end{eqnarray}
where $\Theta\left(x\right)$ is the Heaviside function and $\Phi_{\alpha \mathbf{k}}\left(\mathbf{r}\right)$ is the normal-state eigenfunction of an electron in the $\alpha$th band, $\left[\mathcal{H}_{0}+U_{\alpha }\left(\mathbf{r}\right)\right]
\Phi_{\alpha \mathbf{k}}\left(\mathbf{r}\right)=\tilde{\varepsilon}_{\alpha }\left(\mathbf{k}\right)
\Phi_{\alpha \mathbf{k}}\left(\mathbf{r}\right)$.

In the analogy with the single band case \cite{Ref ketterson}, we obtain on the basis of Eqs. (\ref{eq12}), (\ref{eq13}) the system of Ginzburg-Landau equations for superconductivity gaps in  a two-band system ($\mathbf{A}=0$),
\begin{eqnarray}\label{eq15}
\Delta_{\alpha }(\mathbf{r})&=&-\sum_{\alpha '}w_{\alpha \alpha'}\sqrt{\frac{\rho_{\alpha '}}{\rho_{\alpha }}}\biggl[g(T)-\nu
\left|\Delta_{\alpha '}(\mathbf{r})\right|^{2} \nonumber \\
&+&\sum_{i=1}^{3}\beta_{\alpha 'i}\nabla^{2}_{i}\biggr]\Delta_{\alpha '}(\mathbf{r}) \, ,
\end{eqnarray}
where the intra- and interband interactions are non-zero in the energy layer with the width $2\hbar\omega_{D}$ near the Fermi level, $\varrho_{\alpha }$ is the density of electron states per one spin direction in the $\alpha$th band at the Fermi level, and
\begin{eqnarray}\label{eq16}
w_{\alpha \alpha'}=w_{\alpha '\alpha}=W_{\alpha \alpha'}\sqrt{\rho_{\alpha '}\rho_{\alpha }} \, ,
\end{eqnarray}
\begin{eqnarray}\label{eq17}
g(T)=\ln\biggl({\frac{1.13\hbar\omega_{D}}{k_{B}T}}\biggr) \, ,
\end{eqnarray}
\begin{eqnarray}\label{eq18}
\beta_{\alpha i}=\frac{7\zeta(3)\hbar^{2}v^{2}_{F\alpha i}}{16(\pi k_{B}T_{c})^{2}} \, ,
\end{eqnarray}
\begin{eqnarray}\label{eq19}
\nu=\frac{7\zeta(3)}{8(\pi k_{B}T_{c})^{2}} \, .
\end{eqnarray}
The equations (\ref{eq15}) follow also from the minima conditions of the free energy functional suggested in \cite{Ref Zhi}.

The superconducting transition temperature $T_{c}$ has been determined by the equation
\begin{eqnarray}\label{eq20}
\bigl(1+w_{11}g(T_{c})\bigr)\bigl(1+w_{22}g(T_{c})\bigr)-w_{12}^{2}g^{2}(T_{c})=0 \, .
\end{eqnarray}
In general, Eq. (\ref{eq20}) has two solutions
\begin{eqnarray}\label{eq20a}
&&k_{B}T^{\pm}_{c}=1.13\hbar\omega_{D} \nonumber \\
&\times&\exp\left (\frac{w_{11}+w_{22}\pm\sqrt{\left(w_{11}-w_{22}\right)^{2}+4w^{2}_{12}}}
{2\left(w_{11}w_{22}-w^{2}_{12}\right)}\right ) \, .
\end{eqnarray}
The higher temperature $T^{+}_{c}>T^{-}_{c}$ increases as the interband interaction constant $|w_{12}|$ increases. Simultaneously the temperature $T^{-}_{c}$ decreases, and it disappears as the difference $w_{11}w_{22}-w^{2}_{12}$ approaches $0+$. The temperature region $T>T^{+}_{c}$ corresponds to the normal phase. In the domain $T^{+}_{c}>T>T^{-}_{c}$ there exists a stable superconducting state. If $T<T^{-}_{c}$, a metastable phase (or at least the saddle points of the free energy as a function of non-equilibrium gap order parameters) appears besides the stable superconducting phase \cite{Ref soda}-\cite{Ref ord2}. Consequently, the phase transition into the stable superconducting state takes place at $T=T^{+}_{c}\equiv T_{c}$.

If the interband pairing is absent, $w_{12}=0$, the quantities $T^{\pm}_{c}$ transform into the temperatures of autonomous superconducting phase transitions in the independent bands:
\begin{eqnarray}\label{eq20b}
k_{B}T^{\pm}_{c}=k_{B}T_{c1,2}=1.13\hbar\omega_{D}\exp\left(\frac{1}{w_{11,22}}\right) \, ,
\end{eqnarray}
with $w_{\alpha \alpha}<0$ and $|w_{11}|>|w_{22}|$, i.e. $T_{c1}>T_{c2}$.

In the isotropic situation (the case of the spherical Fermi surface) the squared modulus of the Fermi velocity equals $v^{2}_{F\alpha}=3v^{2}_{F\alpha i}, \,\, i=1,2,3$. Correspondingly, the system of equations (\ref{eq15}) reads
\begin{eqnarray}\label{eq21}
\Delta_{\alpha }(\mathbf{r})&=&-\sum_{\alpha '}w_{\alpha \alpha'}\sqrt{\frac{\rho_{\alpha '}}{\rho_{\alpha }}}\biggl[g(T)-\nu
\left|\Delta_{\alpha '}(\mathbf{r})\right|^{2} \nonumber \\
&+&\beta_{\alpha '}\nabla^{2}\biggr]\Delta_{\alpha '}(\mathbf{r}) \,
\end{eqnarray}
with
\begin{eqnarray}\label{eq22}
\beta_{\alpha }=\frac{7\zeta(3)\hbar^{2}v^{2}_{F\alpha}}{48(\pi k_{B}T_{c})^{2}} \, .
\end{eqnarray}
On the basis of Eqs. (\ref{eq21}) we are going to consider the peculiarities of coherency in the present model.

\section{Critical and non-critical coherence lengths}

We will find now the coherence (correlation) lengths $\xi$ which characterize the spatial variation of small fluctuations of superconducting gaps. By introducing small deviations $\eta_{\alpha }(\mathbf{r})$ from the bulk values of gaps we write
\begin{eqnarray}\label{eq22a}
\Delta_{1,2}(\mathbf{r})=\Delta^{\infty}_{1,2}+\eta_{1,2}(\mathbf{r}) \, .
\end{eqnarray}
In the normal phase $\Delta^{\infty}_{1,2}=0$.
One can take $\Delta_{1,2}(\mathbf{r})$ to be real. Then for the small $\eta_{\alpha }(\mathbf{r})$ we obtain on the basis of Eqs. (\ref{eq21}) the following linearized equations:
\begin{eqnarray}\label{eq22b}
\eta_{\alpha }(\mathbf{r})=-\sum_{\alpha '}w_{\alpha \alpha'}\sqrt{\frac{\rho_{\alpha '}}{\rho_{\alpha }}}\biggl[\tilde{g}_{\alpha '}(T) +\beta_{\alpha '}\nabla^{2}\biggr]\eta_{\alpha '}(\mathbf{r}) \,
\end{eqnarray}
with
\begin{eqnarray}\label{eq22c}
\tilde{g}_{\alpha }(T)=g(T)-3\nu
\left(\Delta^{\infty}_{\alpha }(T)\right)^{2} \,
\end{eqnarray}

\begin{figure}[!ht]
\begin{center}
\resizebox{1.00\columnwidth}{!} {\includegraphics[angle=-90]{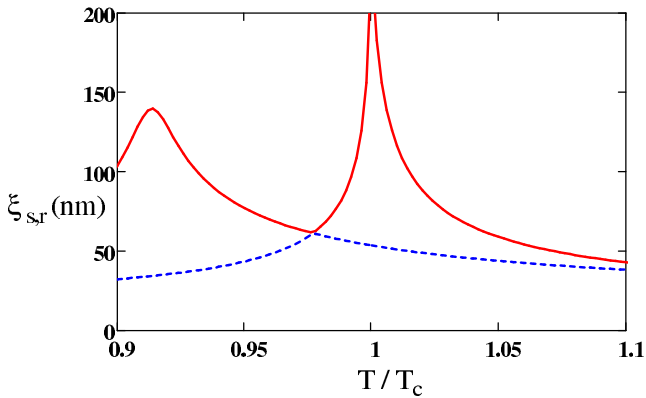}}
\caption{The coherence lengths $\xi_{s}$ (solid line) and $\xi_{r}$ (dashed line) \textit{vs} temperature. Note that the minimum of $\xi_{s}$ and the maximum of $\xi_{r}$ are actually separated by a small gap. Parameters: $W_{11}=-0.3 \, eV\cdot cell$, $W_{22}=-0.57 \, eV\cdot cell$, $|W_{12}|=0.0001 \, eV\cdot cell$, $\rho_{1}=1 \, (eV\cdot cell)^{-1}$, $\rho_{2}=0.5 \, (eV\cdot cell)^{-1}$, $\hbar\omega_{D}=0.07 \, eV$, $v_{F1}=4\times 10^{5} \, m/s$, $v_{F2}=5\times 10^{5} \, m/s$. Characteristic temperatures: $T_{c}=T_{c}^{+}\approx T_{c1}=30.4 \, K $, $T_{c}^{-}\approx T_{c2}=27.6 \, K$. } \label{f1}
\end{center}
\end{figure}

\begin{figure}[!ht]
\begin{center}
\resizebox{1.00\columnwidth}{!} {\includegraphics[angle=-90]{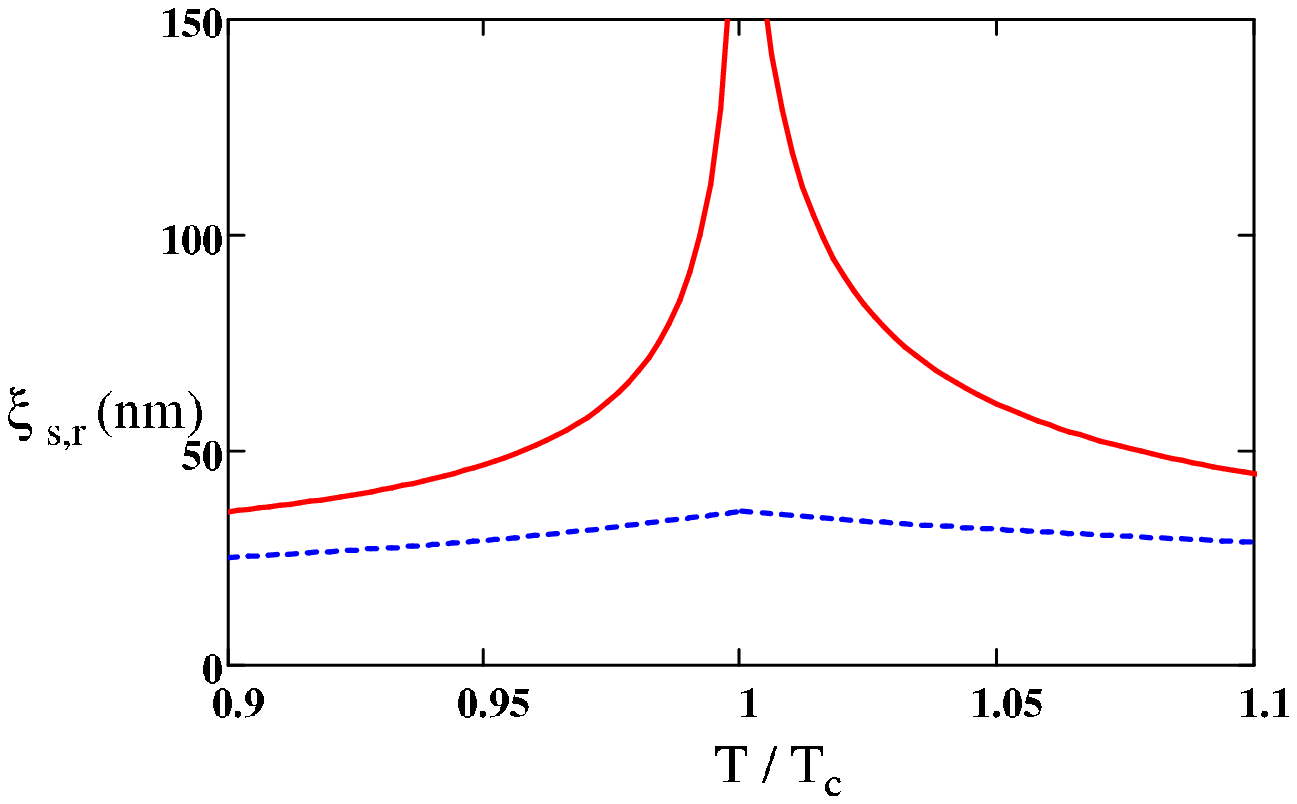}}
\caption{The coherence lengths $\xi_{s}$ (solid line) and $\xi_{r}$ (dashed line) \textit{vs} temperature. Parameters: $W_{11}=-0.3 \, eV\cdot cell$, $W_{22}=-0.57 \, eV\cdot cell$, $|W_{12}|=0.009 \, eV\cdot cell$, $\rho_{1}=1 \, (eV\cdot cell)^{-1}$, $\rho_{2}=0.5 \, (eV\cdot cell)^{-1}$, $\hbar\omega_{D}=0.07 \, eV$, $v_{F1}=4\times 10^{5} \, m/s$, $v_{F2}=5\times 10^{5} \, m/s$. Characteristic temperatures: $T_{c}=T_{c}^{+}=31.6 \, K$, $T_{c1}=30.4 \, K $, $T_{c}^{-}=26.5 \, K,$ $T_{c2}=27.6 \, K$.}
\label{f2}
\end{center}
\end{figure}

\begin{figure}[!ht]
\begin{center}
\resizebox{1.00\columnwidth}{!} {\includegraphics[angle=-90]{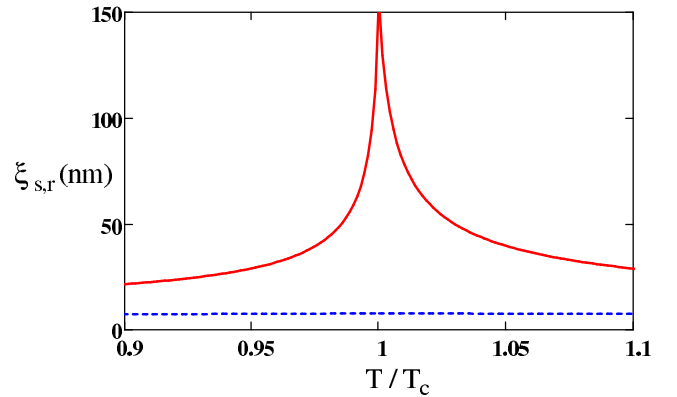}}
\caption{The coherence lengths $\xi_{s}$ (solid line) and $\xi_{r}$ (dashed line) \textit{vs} temperature. Parameters: $W_{11}=-0.3 \, eV\cdot cell$, $W_{22}=-0.57 \, eV\cdot cell$, $|W_{12}|=0.08 \, eV\cdot cell$, $\rho_{1}=1 \, (eV\cdot cell)^{-1}$, $\rho_{2}=0.5 \, (eV\cdot cell)^{-1}$, $\hbar\omega_{D}=0.07 \, eV$, $v_{F1}=4\times 10^{5} \, m/s$, $v_{F2}=5\times 10^{5} \, m/s$. Characteristic temperatures: $T_{c}=T_{c}^{+}=50.9 \, K$, $T_{c1}=30.4 \, K $, $T_{c}^{-}=12.6 \, K$, $T_{c2}=27.6 \, K$.}
\label{f3}
\end{center}
\end{figure}

We seek for the solutions of equations (\ref{eq22b}) in the form
\begin{eqnarray}\label{eq23}
\eta_{1,2}(\mathbf{r})\sim\mathrm{exp}\biggl(-\frac{\sum_{i=1}^{3}x_{i}}{\sqrt{3}\xi}\biggr) \, ,
\end{eqnarray}
where $\xi$ represents the length scales.
The corresponding substitution into Eqs. (\ref{eq22b}) yields
\begin{eqnarray}\label{eq24}
\biggl(1+w_{11}\tilde{g}_{1}(T)+w_{11}\beta_{1}\xi^{-2}\biggr)\eta_{1}(\mathbf{r}) \nonumber \\
+\biggl(w_{12}\sqrt{\frac{\rho_{2}}{\rho_{1}}}\tilde{g}_{2}(T)
+w_{12}\sqrt{\frac{\rho_{2}}{\rho_{1}}}\beta_{2}\xi^{-2}\biggr)\eta_{2}(\mathbf{r})&=&0\nonumber\\
\biggl(w_{21}\sqrt{\frac{\rho_{1}}{\rho_{2}}}\tilde{g}_{1}(T)+w_{21}\sqrt{\frac{\rho_{1}}{\rho_{2}}}\beta_{1}\xi^{-2}\biggr)\eta_{1}(\mathbf{r}) \nonumber \\
+\biggl(1+w_{22}\tilde{g}_{2}(T)
+w_{22}\beta_{2}\xi^{-2}\biggr)\eta_{2}(\mathbf{r})&=&0 \, .
\end{eqnarray}
The system of equations (\ref{eq24}) has the non-zero solutions $\eta_{1,2}(\mathbf{r})$ if the determinant of this linear homogeneous system equals to zero.
The latter condition leads to a bi-quadratic equation for the characteristic lengths $\xi$
\begin{eqnarray}\label{eq26}
K\left(T\right)\xi^{4} - G\left(T\right)\xi^{2} + \gamma = 0 \,
\end{eqnarray}
where
\begin{eqnarray}\label{eq27}
G\left(T\right)&=&w^{2}_{12}\left[\tilde{g}_{1}(T)\beta_{2}+\tilde{g}_{2}(T)\beta_{1}\right] \nonumber\\ &-&\left[1+w_{11}\tilde{g}_{1}(T)
\right]w_{22}\beta_{2} \nonumber\\
&-&\left[1+w_{22}\tilde{g}_{2}(T)\right]w_{11}\beta_{1} \, ,
\end{eqnarray}
\begin{eqnarray}\label{eq28}
K\left(T\right)&=&\left[1+w_{11}\tilde{g}_{1}(T)\right]\left[1+w_{22}\tilde{g}_{2}(T)\right] \nonumber\\
&-&w_{12}^{2}\tilde{g}_{1}(T)\tilde{g}_{2}(T) \, ,
\end{eqnarray}
\begin{eqnarray}\label{eq29}
\gamma=(w_{11}w_{22}-w^{2}_{12})\beta_{1}\beta_{2} \, .
\end{eqnarray}
Solving Eq. (\ref{eq26}) we obtain the expressions for two squared characteristic lengths
\begin{eqnarray}\label{eq30}
\xi_{s,r}^{2}\left(T\right)=\frac{G\left(T\right)
\pm\sqrt{G^{2}\left(T\right)-4K\left(T\right)\gamma}}{2K\left(T\right)} \, .
\end{eqnarray}
One can observe from Eq. (\ref{eq30}) that the quantities $\xi_{s}\left(T\right)$ (soft or critical coherence length) and $\xi_{r}\left(T\right)$ (rigid or non-critical coherence length) reveal substantially different temperature behavior near $T_{c}$ \footnote{Note that $K\left(T_{c}\right)=0$ according to Eq. (\ref{eq20}).}. The characteristic length $\xi_{s}\left(T\right)$ behaves critically diverging at the phase transition point: $\xi_{s}\left(T_{c}\right)=\infty$.
At the same time $\xi_{r}\left(T\right)$ remains finite, $\xi_{r}\left(T_{c}\right)=\sqrt{\gamma /G\left(T_{c}\right)}$, and its temperature dependence is weaker.
The functions $\xi_{s,r}\left(T\right)$ have been depicted in Figs. \ref{f1}-\ref{f3} with increasing interband interaction constant $|W_{12}|$. In the temperature domain $T<T_{c}$ used in Figs. \ref{f1}-\ref{f3} the deviation of the solutions of the homogeneous part of approximate equations Eq. (\ref{eq21}) from the solutions of exact equations Eq. (\ref{eq12}) is very small, less than 5\%.


In Fig. \ref{f1}, the maximum of $\xi_{s}(T)$ near $T^{-}_{c}\approx T_{c2}$ reflects the "memory" about the lower autonomous phase transition which takes place if interband interaction is absent. The increase of $|W_{12}|$ suppresses this maximum and finally $\xi_{s}(T)$ decreases monotonically as temperature $T<T_{c}$ decreases, cf. Figs. \ref{f1} and \ref{f2}. At the same time the temperature dependence of $\xi_{r}$ becomes weaker as one can observe from Figs. \ref{f1}-\ref{f3}. The further increase of interband coupling does not introduce any qualitative changes into the temperature behaviour of coherence lengths.

The length scales found are related to the critical and non-critical fluctuations which appear as linear combinations of the fluctuations of band superconductivity order parameters, see also \cite{Ref ord1}, \cite{Ref kristoffel}.

One can find by using the expansion in powers of $(T-T_{c})/T_{c}$ that the following expressions approximate the temperature dependence of the critical coherence length near the phase transition point:
\begin{equation}\label{eq31}
\xi_{s}(T)=
\left\{%
\begin{array}{cc}
  \xi_{s}^{0}\sqrt{\frac{T_{c}}{T-T_{c}}} \, , & T>T_{c} \\
  \\
  \xi_{s}^{0}\sqrt{\frac{T_{c}}{2(T_{c}-T)}} \, , & T<T_{c} \\
\end{array}%
\right.\, ,
\end{equation}
where
\begin{eqnarray}\label{eq32}
\xi_{s}^{0} =\sqrt{\frac{G\left(T_{c}\right)}{-w_{11}-w_{22}-2(w_{11}w_{22}-w^{2}_{12})g(T_{c})}
} \, .
\end{eqnarray}
The coefficient $\xi_{s}^{0}$ coincides with the length found in \cite{Ref kogan} as a single coherence length in the rigorously limited Ginzburg-Landau scheme in a two-band superconductor.

\section{Coherence length-scales and microscopic length-scales}

The gradient expansion and its cut-off in the Ginzburg-Landau equations (\ref{eq21}) is justified if
\begin{eqnarray}\label{eq33}
\frac{\xi_{s,r}^{2}(T)}{\beta_{1,2}} \gg 1 \, ,
\end{eqnarray}
i.e. the length scales $\xi_{s,r}(T)$ are large enough compared to the microscopic lengths $\sqrt{\beta_{1,2}}$.

\begin{figure}[!ht]
\begin{center}
\resizebox{0.90\columnwidth}{!} {\includegraphics[angle=-90]{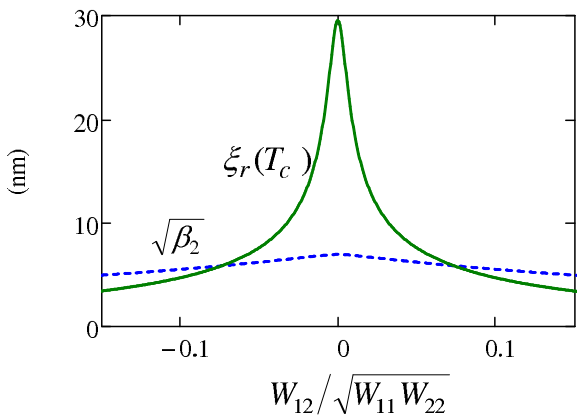}}
\caption{The dependence of the rigid coherence length $\xi_{r}(T_{c})$ (solid line) and the microscopic length $\sqrt{\beta_{2}}$ (dashed line) on the interband interaction constant $W_{12}$. The intraband interactions are fixed: $W_{11}=-0.10 \, eV\cdot cell$, $W_{22}=-0.78 \, eV\cdot cell$. Parameters: $\rho_{1}=4\, (eV\cdot cell)^{-1}$, $\rho_{2}=0.5 \, (eV\cdot cell)^{-1}$, $\hbar\omega_{D}=0.07 \, eV$, $v_{F1}=2.5\times 10^{5} \, m/s$, $v_{F2}=5\times 10^{5} \, m/s$.}
\label{f4}
\end{center}
\end{figure}

It is relatively easy to satisfy the inequality (\ref{eq33}) for $\xi_{s}(T)$.
However, the situation is different in the case the non-critical coherence length $\xi_{r}(T)$. In Fig. \ref{f4} the comparison of the dependencies of $\xi_{r}(T_{c})$ and $\sqrt{\beta_{2}}$ on interband interaction constant is demonstrated (we have chosen $v_{F2}>v_{F1}$, i.e. $\beta_{2}>\beta_{1}$). It is seen that in the region of weak interband coupling the condition (\ref{eq33}) can be satisfied for $\xi_{r}$. At that the smaller the difference between the temperatures of autonomous phase transitions $T_{c1}$ and $T_{c2}$ in the absence of interband interaction, the higher and sharper in Fig. \ref{f4} the maximum of $\xi_{r}(T_{c})$ \textit{vs} $W_{12}$ is. For larger $|W_{12}|$ the condition (\ref{eq33}) for $\xi_{r}$ becomes violated. Consequently, in this region of parameters the higher terms in the gradient expansion should be taken into account for the treatment of the non-critical length scale of coherency. Recently the systematic extension of the Ginzburg-Landau scheme has been suggested in \cite{Ref shanenko1}, \cite{Ref shanenko2}.

\section{Conclusions}

To conclude, we determined two coherence lengths in a two-gap superconductor with intra- and interband couplings. These length scales are not related to the concrete bands involved in the formation of the superconducting ordering in a system with interband pair transfer interaction. One of these lengths as a function of temperature deverges near the phase transition point, the other one is non-critical. The non-monotonic temperature dependence of coherence lengths appears in the superconducting phase if the interband coupling is sufficiently weak. We suggest that the appearance of the non-critical coherence length is a substantial feature of two-gap superconductivity.

\begin{acknowledgements}
This research was supported by the European Union through the European Regional Development Fund (Centre of Excellence "Mesosystems: Theory and Applications", TK114). We acknowledge the support by the Estonian Science Foundation, Grant No 7296.
\end{acknowledgements}


\begin{thebibliography}{}
%
%

\bibitem{Ref smw}
\nimekirja{\Nauthor{H.,}{Suhl} \Nauthor{B.T.,}{Matthias} \Nauthor{L. P.}{Walker}}{Phys. Rev.\\ Lett.}{3}{552}{1959}
\bibitem{Ref m}
\nimekirja{\Nauthor{V. A.}{Moskalenko}}{Fiz. Met. Metalloved.}{8}{503}{1959}
\bibitem{Ref kondo}
\nimekirja{\Nauthor{J.}{Kondo}}{Progr. Theor. Phys.}{29}{1}{1963}
\bibitem{Ref kw}
\nimekirja{\Nauthor{V. Z.,}{Kresin} \Nauthor{S. A.}{Wolf}}{Phys. Rev. B}{46}{6458}{1992}
\bibitem{Ref kko}
\nimekirja{\Nauthor{N.,}{Kristoffel} \Nauthor{P.,}{Konsin} \Nauthor{T.}{\"{O}rd}}{Riv. Nuovo Cimen-\\to}{19}{8}{1994}
\bibitem{Ref biankoni1}
\nimekirja{\Nauthor{A.}{Bianconi}}{J. Superconductivity}{18}{625}{2005}
\bibitem{Ref biankoni2}
\nimekirja{\Nauthor{R.,}{Caivano} \Nauthor{M.,}{Fratini} \Nauthor{N.,}{Poccia} \Nauthor{A.,}{Ricci} \Nauthor{A.,}{Puri} \Nauthor{Z.-A.,}{Ren} \Nauthor{X.-L.,}{Dong} \Nauthor{J.,}{Yang} \Nauthor{W.,}{Lu} \Nauthor{Z.-X.,}{Zhao} \Nauthor{ L.,}{Barba} \Nauthor{A.}{Bianconi}}{Supercond. Sci. Technol.}{22}{014004}{2009}
\bibitem{Ref oxygen}
\nimekirja{\Nauthor{N.,}{Poccia} \Nauthor{M.,}{Fratini} \Nauthor{A.,}{Ricci} \Nauthor{G.,}{Campi} \Nauthor{ L.,}{Barba} \Nauthor{ A.,}{Vittorini-Orgeas} \Nauthor{G.,}{Bianconi} \Nauthor{G.,}{Aeppli} \Nauthor{A.}{Bianconi}}{Nature Materials}{10}{733}{2011}
\bibitem{Ref babaev}
\nimekirja{\Nauthor{J.,}{Carlstr\"{o}m} \Nauthor{E.,}{Babaev} \Nauthor{M.}{Speight}}{Phys. Rev. B}{83}{174509}{2011}
\bibitem{Ref poluektov}
\nimekirja{\Nauthor{Y. M.,}{Poluektov} \Nauthor{V. V.}{Krasilnikov}}{Fizika Nizkhik Temperatur}{15}{1251}{1989}
\bibitem{Ref konsin}
\nimekirja{\Nauthor{P.}{Konsin}}{Phys. Stat. Sol. (b)}{189}{185}{1995}
\bibitem{Ref kristoffel}
\nimekirja{\Nauthor{N.,}{Kristoffel} \Nauthor{T.,}{\"{O}rd} \Nauthor{P.}{Rubin}}{Supercond. Sci. Technol.}{22}{014006}{2009}
\bibitem{Ref ketterson}
\nimekirja{\Nauthor{J. B.,}{Ketterson} \Nauthor{S. N.$\!$}{Song}}{Superconductivity$\!\!$}{}{497 p. Cambridge University Press, Cambridge}{1999}
\bibitem{Ref soda}
\nimekirja{\Nauthor{T.,}{Soda} \Nauthor{K. Y.}{Wada}}{Progr. Theor. Phys.}{36}{1111\\}{1966}
\bibitem{Ref ord1}
\nimekirja{\Nauthor{T.,}{\"{O}rd} \Nauthor{K.,}{R\"{a}go} \Nauthor{A.}{Vargunin}}{J. Supercond. Novel Magn.}{22}{85}{2009}
\bibitem{Ref ord2}
\nimekirja{\Nauthor{T.,}{\"{O}rd} \Nauthor{K.,}{R\"{a}go} \Nauthor{A.}{Vargunin}}{In: Physical Proper-\\ties of Nanosystems, Ed. J. Bon\v{c}a, S. Kruchinin, NATO Science for Peace and Security Series B: Physics and Biophysics}{}
{p.177, Springer, Dordrecht}{2011}
\bibitem{Ref Zhi}
\nimekirja{\Nauthor{M. E.,}{Zhitomirsky} \Nauthor{V.-H.}{Dao}}{Phys. Rev. B}{69}{\\054508}{2004}
\bibitem{Ref kogan}
\nimekirja{\Nauthor{V. G.,}{Kogan} \Nauthor{J.}{Schmalian}}{Phys. Rev. B}{83}{\\054515}{2011}
\bibitem{Ref shanenko1}
\nimekirja{\Nauthor{A. A.,}{Shanenko} \Nauthor{M. V.,}{Milo\v{s}evi\'{c}} \Nauthor{F. M.}{Peeters}}
{\\Phys. Rev. Lett}{106}{047005}{2011}
\bibitem{Ref shanenko2}
\nimekirja{\Nauthor{L.,}{Komendov\'{a}} \Nauthor{M. V.,}{Milo\v{s}evi\'{c}} \Nauthor{A. A.,}{Shanenko} \Nauthor{F. M.}{Peeters}}{Phys. Rev. B}{84}{064522}{2011}
\end{thebibliography}

\newcommand{\Nauthor}[2]{#2,$\;$#1$\;$}
\newcommand{\nimekirja}[5]{#1:$\,\;${#2}$\;$\textbf{#3},$\;${#4}$\;${(#5)}.}

\end{document}